# On atomic structure of Guinier-Preston Zone in Mg-Zn-Gd alloy

Xin-Fu Gu[1,2], Tadashi Furuhara[2], Takanori Kiguchi[2], Toyohiko J. Konno[2], Leng Chen[1], Ping Yang[1]

1. School of Materials Science and Engineering, University of Science and Technology Beijing, Beijing, 100083, China
2. Institute for Materials Research, Tohoku University, Sendai, 980-8577, Japan

**Abstract**

The atomic structure of Guinier–Preston (G.P.) zone in an Mg-Zn-Gd alloy has been determined by combining atomic resolution high angle annular dark field (HAADF) scanning transmission electron microscopy (STEM) and first-principles calculation. The G.P. Zone consists of three atomic layers paralleling to (0001) plane of Mg matrix, where two outer layers enrich of Gd atoms while the middle layer enriches of Zn atoms. Moreover, it is found that most reported structures in literature are unstable and will converge to present result after atomic relaxation, even though the atomic species and positions in the middle layer vary in those models.

Magnesium alloy containing rare earth elements usually exhibits remarkable age-hardening effect and good creep resistance [1-3], due to formation of coherent/semi-coherent metastable phases during ageing process [4]. Other elements are often added to enhance the hardenability and replace part of expensive rare earth elements [5, 6]. The addition of Zn to Mg-Gd alloy will significantly enhance the ageing response [7], where densely distributed GP zones [8-10] and/or long-period stacking ordered (LPSO) structures will form [6, 11]. Contrary to intensive work on LPSO structure [11-15], the atomic structure of G.P. Zone has not been unambiguously determined. Nishijima *et al.* [8] first proposed the atomic structure of G.P. zone based on a high resolution TEM and high-angle annular detector dark field scanning transmission electron microscopy (HAADF-STEM) study in $Mg_{97}Zn_1Gd_2$ (at.%, default) alloy. They found G.P. zone is fully coherent with matrix and has a hexagonal structure with its base vectors along <1-100>$_\alpha$ in matrix. The lattice parameters are $\sqrt{3}a_\alpha$ and $c_\alpha$, where $a_\alpha$ and $c_\alpha$ are the lattice parameters of Mg matrix. The G.P. Zone consists of three atomic layers paralleling to (0001)$_\alpha$ plane in Mg matrix, where Gd and/or Zn atoms occupy ordered positions (0, 0, 0) in the outer two layers while the middle layer is pure Mg. However, Nie *et al.* [9] pointed out that the middle layer in the G.P. zone (namely γ″ in their work) should enrich of Zn instead of pure Mg based on their HAADF-STEM and three dimension atom probe (3DAP) study in Mg$_{98.4}$Gd$_1$Zn$_{0.4}$Zr$_{0.2}$ alloy. They also noted that there is a negative misfit between the G.P. zone and the Mg matrix along [0001]$_{G.P.}$ // [0001]$_\alpha$. Very recently, *Li et al.* [16] proposed a new atomic structure for G.P. Zone in a HAADF-STEM study of aged Mg$_{96.5}$Zn$_1$Gd$_{2.5}$ alloy, where the outer two layers are same as previous models, but atomic positions in the middle layer are different. They found that the atoms in middle layer are shifted comparing to the matrix structure from a [10-10]$_\alpha$ zone axis view. Therefore, the atomic positions in the middle layer are proposed to shift a distance of $a_\alpha$/4 along [11-20]$_\alpha$ to keep the atomic structure along [11-20]$_\alpha$ invariant but the atomic structure along [10-10]$_\alpha$ in agree with the observation. However, this translation will make the projections of the atomic structure along three <10-10>$_\alpha$ or <11-20>$_\alpha$ directions inequivalent (Refer to different projections of atomic structure in Figure S5 in supplementary material), which conflicts with other experimental observations. In addition, Zhu *et al.* [17] have obtained similar observation in an Ag modified Mg-Zn-Y alloy where the atomic positions in middle layer are shifted when viewing at not only <10-10>$_\alpha$, but also <11-20>$_\alpha$ directions. However, the precipitate in their study enriches of Ag instead of Zn. Therefore, a detailed study on the atomic structure of G.P. zone in Mg-Zn-Gd alloy is needed.

As shown above, HAADF-STEM plays an important role in determining atomic structure, and the advent of Cs-corrected STEM has made it possible to directly compare the proposed crystal structures with experimental observations [17-20]. In addition, the combination of first-principles calculation with Cs-corrected STEM has become an inevitable method to study the atomic structures [13, 17, 19]. In this work, we will determine the atomic structure of the G.P. zone in an Mg-Zn-Gd alloy by atomic resolution HAADF-STEM, and will further compare the structure with other models in literature by first-principles calculation.

The as-casted $Mg_{97}Zn_{1.8}Gd_{1.2}$ alloy was solution treated at 520°C for 2 hours, and then aged at 280°C for 1 hour. TEM specimens of $\phi 3 \times 0.5$ mm were cut from the heat treated specimens by using electro-discharge machine. The discs were then grinded to about 60 μm and thinned by a dimpling grinder (Model 2000, Fischione) followed by a low-energy ion milling from 4 kV to 500 V with an incidence angle of 4° (PIPS model 691; Gatan Inc.). The HAADF-STEM observation was carried out with Titan G2 60-300 (300kV, FEI), and the collection semi-angle of angular detector was 50.5-200 mrad. The 3DAP data was taken by CAMECA Leap 4000 at 35K. A pulse fraction of 20% and a pulse frequency of 200 kHz were used. The first-principles calculation was performed with Cambridge Serial Total Energy Package (CASTEP) code which implements the plane wave pseudo-potential method of the density functional theory [21]. The Perdew-Bruke-Ernzerhof (PBE) version of generalized gradient approximation (GGA) was used to describe electron exchange correlation functional [22]. The lattice constants and atomic positions were optimized with Broyden-Fletcher-Goldfarb-Shanno (BFGS) method [23], where the tolerance for energy change, maximum force and maximum displacement in fully relaxed state was set to be $5.0 \times 10^{-6}$ eV/atom, 0.01 eV/Å, $5.0 \times 10^{-4}$ Å, respectively.

Figure 1 shows a low mag view of the precipitates along $[11\text{-}20]_\alpha$ and $[0001]_\alpha$ zone axis by HAADF-STEM. The white contrasts in the image are precipitates due to enrichment of Gd and/or Zn according to Z-contrast principle, because the atomic number (Z) for three elements in this study are 12 for Mg, 30 for Zn and 64 for Gd. Later we will show the precipitates are G.P. zones. In Figure 1(a), densely distributed precipitates are edge-on, and the habit plane is parallel to $(0001)_\alpha$. From a $[0001]_\alpha$ view in Figure 1(b), the plate-like G.P. Zone has a hexagonal shape with its edge along $<10\text{-}10>_\alpha$. The compositions of these precipitates are measured by 3DAP, and the result is shown in Figure 2. Similar to HAADF-STEM, plate-like precipitates could also be observed, and they are all synchronized with Zn and Gd as seen from the element distribution in Figure 2(a). Figure 2(b) shows composition profile across a precipitate. The Zn and Gd contents are quite high and around 15 at%, and the Zn content is a little higher than Gd. This result is similar to the result obtained by Nie *et al* in an $Mg_{98.4}Gd_1Zn_{0.4}Zr_{0.2}$ alloy [9]. It should be noted that the precipitate is too thin to measure the composition accurately due to the aberrations in 3DAP test.

In order to determine the atomic structure of precipitates, high resolution observation is carried out from three zone axes, i.e. $[11\text{-}20]_\alpha$, $[10\text{-}10]_\alpha$ and $[0001]_\alpha$, by HAADF-STEM, and the result is shown in Figure 3. Figure 3(a-c) are corresponding diffraction patterns. Except for the diffraction spots from Mg matrix, extra diffraction strike or spots are found along $(0002)_\alpha$ at positions of $\{1\text{-}100\}_\alpha$, $\{11\text{-}20\}_\alpha/3$, $2\{11\text{-}20\}_\alpha/3$ and $\{11\text{-}20\}_\alpha$. The extra spots are reported due to ordering effect in the structure of precipitates [8, 9]. Figure 3(d-f) shows the atomic structures from $[11\text{-}20]_\alpha$, $[10\text{-}10]_\alpha$ and $[0001]_\alpha$ zone axes, respectively. The hexagonal-shaped precipitate found in Figure 1 is coherent with the matrix, and consists of three layers paralleling to $(0002)_\alpha$ plane. Therefore, the precipitates in this study are the same as those in previous studies [8, 9, 16], namely, the precipitates are G.P. zones. The intensity profile from row 1, 2 and 3 in Figure 3(e) are shown in Figure 3(g). The outer two layers in G.P. zone exhibit stronger contrast than the middle layer or Mg layer, which means there are much heavier atoms (Gd) or dense atomic rows along the zone axis. Ordering of Gd atoms in outer layer can be directly seen from $[0001]_\alpha$ as in Figure 3(f). The Gd atoms periodically locate at $<10\text{-}10>_\alpha$ positions by substitute corresponding Mg atoms, thus a Gd atom can be found at a distance of every $3a_\alpha$ along $<11\text{-}20>_\alpha$, which could explain the extra spots observed in the diffraction patterns in Figure 3(a-c). Such ordering of Gd atoms are usually deduced from the diffraction patterns [8, 9]. It is the first time to directly observe the atomic structure by HAADF-STEM, though the structure of outer layer in previous studies is consistent with present result. Similar to previous studies [9, 16], the spacing between two Gd layers is measured to be around 3.9 Å, and is smaller than $c_\alpha = 5.21$ Å, indicating a negative strain between the G.P. zone and matrix. As shown before, the atomic position or species in middle layer varies in different models. In this study, the atoms in the middle layer are indicated by yellow circles in Figure 3, which is different from the atomic position of Mg layer in Mg matrix. Particularly, the atomic rows with stronger contrast in this layer are located at the center between four neighboring Gd atoms marked by red circles in both [11-20]$_\alpha$ and [10-10]$_\alpha$ view in Figure 3(d-e). Li *et al.* [16] pointed out this fact only from [10-10]$_\alpha$ zone axis, but failed from [11-20]$_\alpha$ zone axis. In fact, if we carefully examine their image, the atoms are actually located at centering position. According to the contrast in Figure 3(d-f), it is reasonable to deduce that the middle layer should enrich of Zn.

In considering the Zn atoms centring between Gd atoms, the atomic position of Zn is fixed to be (1/2, 1/2 ,1/2), (1/2, 0, 1/2), and (0, 1/2, 1/2). Therefore, the structure of the G.P. zone could be fully determined. The lattice parameters of the hexagonal structure are $a$ = 5.56 Å ( = $\sqrt{3}a_\alpha$ ) and $c$ = 3.9 Å. The atomic positions of Gd atoms locate at (0, 0, 0), Mg locate at (1/3, 1/3 ,0) and (2/3, 2/3, 0), and Zn locate at (1/2, 1/2 ,1/2), (1/2, 0, 1/2), and (0, 1/2, 1/2). Consequently, the nominal composition of G.P. zone is Mg$_2$Zn$_3$Gd$_1$, but the Zn content is larger than measured value by 3DAP. It is possible that the middle layer contains Mg atoms substituting part Zn atoms. The projections of the atomic structure in different directions are shown in Figure 4. Figure 4(a-c) is the rigid model proposed in this study, while Figure 4(b-f) is the model from Nie *et al.* [9]. Present atomic structure can be treated as a modified version from Nie *et al.* [9], where the Zn atoms in their model are shuffled $\sqrt{3}a_\alpha/6$ along <10-10>$_\alpha$ as indicated by the red arrows in Figure 4(f).

The stability of these structure models will be examined next. In the first-principles calculation, a G.P. zone with three atomic layers is embedded in the Mg matrix with a total of 10 (0001)$_\alpha$ layers and the basal size of $\sqrt{3}a_\alpha \times \sqrt{3}a_\alpha$. The orientation relationship between the G.P. zone and matrix is set to be (0001)$_{G.P.}$ // (0001)$_\alpha$ and [2-1-10]$_{G.P.}$ // [1-100]$_\alpha$ according to Figure 3. As a result, 30 atoms are included in our calculation. The initial structure varies depending on the atomic structure of G.P. zone. The detailed initial structures and corresponding relaxed structures are shown in the supplementary material due to space limitation. The relaxed structure of the present model is shown in Figure 4(g-i). The relaxation mainly happens in the outer two layers of G.P. zone, and the atoms move along $[0001]_\alpha$ after relaxation. Specifically, the Gd atoms move inward and result in a spacing between two Gd layers to be 3.52 Å which is close to 3.9 Å measured from Figure 3, and the Mg atoms in the outer layers shrink to a spacing of 4.84 Å. Nevertheless, the relaxed structure is close to initial structure in Figure 1(a-c), especially when view from $[0001]_\alpha$. The relaxed structure is superimposed onto Figure 3(d-f), showing good agreement. Furthermore, we take the model proposed by Nie *et al.* [9] as an input for atomic relaxation, and found that the structure is not stable and will result in the same structure as Figure 4(g-i) after relaxation. Namely, Zn atoms in the middle layer will move about 0.9 Å ( = $\sqrt{3}a_\alpha/6$ ) along $<10\text{-}10>_\alpha$ due to the atomic interactions. Similarly, if we take the model proposed by Li *et al.* [16] as an input, their structure is also not stable. The atoms in the middle layer will also move during relaxation and the final structure is similar to Figure 4(g-i) as shown in Figure S6(a-c). Moreover, we also checked the effect of Zn/Mg ratio in the middle layer on the relaxation structure, and found the relaxed structure is also similar to previous cases (Refer to Figure S6(d-f)). Therefore, despite of different models in literature, they will converge to present model after atomic relaxation.

In summary, a unified atomic structure of G.P. zone in Mg-Zn-Gd alloy has been determined by combining atomic resolution high angle annular dark field (HAADF) scanning transmission electron microscopy (STEM) and first-principles calculation. The hexagonal-shaped G.P. Zone is found to be consisted of three atomic layers paralleling to (0001) plane in Mg matrix, where the outer two layers enrich of Gd atoms while the middle layer enriches of Zn atoms. The atoms in the middle layer are shifted along <10-10> directions with a distance of about 0.9 Å. The proposed atomic structure has been compared with existing models in literature by first-principles calculation. It shows that all of the structure models reported in literatures are not stable, and they will relax to a structure close to present model.

**Acknowledgement**

This work was supported by a Grant-in-Aid for Scientific Research on Innovative Areas, "Synchronized Long-Period Stacking Ordered Structure", from the Ministry of Education, Culture, Sports, Science and Technology, Japan (No.23109001). The Mg alloy samples used in the present study were supplied by Kumamoto University. Special thanks to Mr. Y. Hayasaka for technical support at Electron Microscopy Center, Tohoku University, Japan.

Figures

Figure 1. Low-mag view of G.P. zone from different zone axes by HAADF-STEM. a) [11-20]$_\alpha$, b) [0001]$_\alpha$. The white contrast in the image is G.P. Zone due to enrichment of atoms with large atomic number.

Figure 2. Three dimensional distribution of solute elements in aged sample taken by 3DAP. a) Element mapping of Zn (blue dot) and Gd (green dot) atoms. Dark blue or green indicates the locations of G.P. zones due to enrichment of solute atoms. b) Concentration profile across a G.P. zone and insert figure showing corresponding G.P. zone.

Figure 3. High-mag view of G.P. zone from different zone axes by HAADF-STEM. a-c) Diffraction patterns taken from [11-20]$_\alpha$, [10-10]$_\alpha$ and [0001]$_\alpha$, respectively. d) Atomic structure viewed from [11-20]$_\alpha$, e) Atomic structure viewed from [10-10]$_\alpha$, f) Atomic structure viewed from [0001]$_\alpha$. g) Intensity profile from row 1-3 indicated in e). In addition, the relaxed atomic structure is superimposed on d-f).

Figure 4. Atomic structures of G.P. zone in different models. a-c) Rigid atomic structure proposed in this study, d-f) Atomic structure model by Nie *et al.* [9], g-h) Relaxed atomic structure for a-c), respectively, by first-principles calculation. Figure a), d) and g) are viewed from [11-20]$_\alpha$, Figure b), e) and h) viewed from [10-10]$_\alpha$, Figure c), f), and i) are viewed from [0001]$_\alpha$.

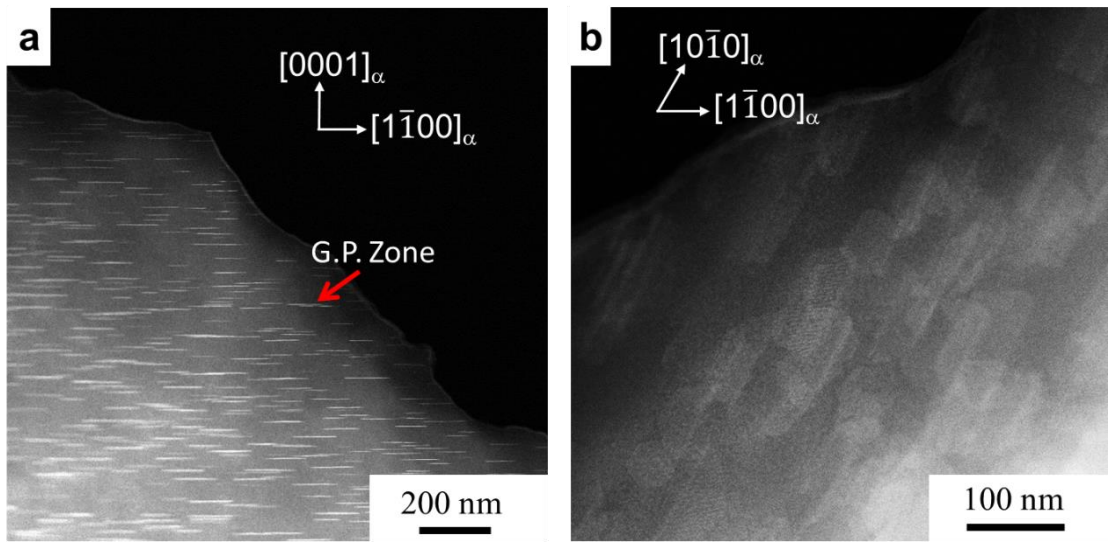

Figure 1. Low-mag view of G.P. zone from different zone axes by HAADF-STEM. a) [11-20]$_α$, b) [0001]$_α$. The white contrast in the image is G.P. Zone due to enrichment of atoms with large atomic number.

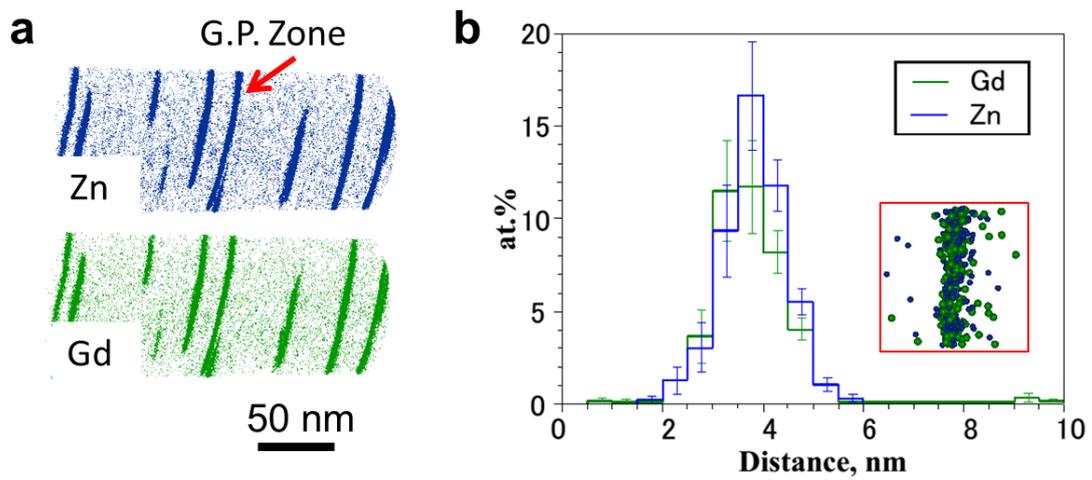

Figure 2. Three dimensional distribution of solute elements in aged sample taken by 3DAP. a) Element mapping of Zn (blue dot) and Gd (green dot) atoms. Dark blue or green indicates the locations of G.P. zones due to enrichment of solute atoms. b) Concentration profile across a G.P. zone and insert figure showing corresponding G.P. zone.

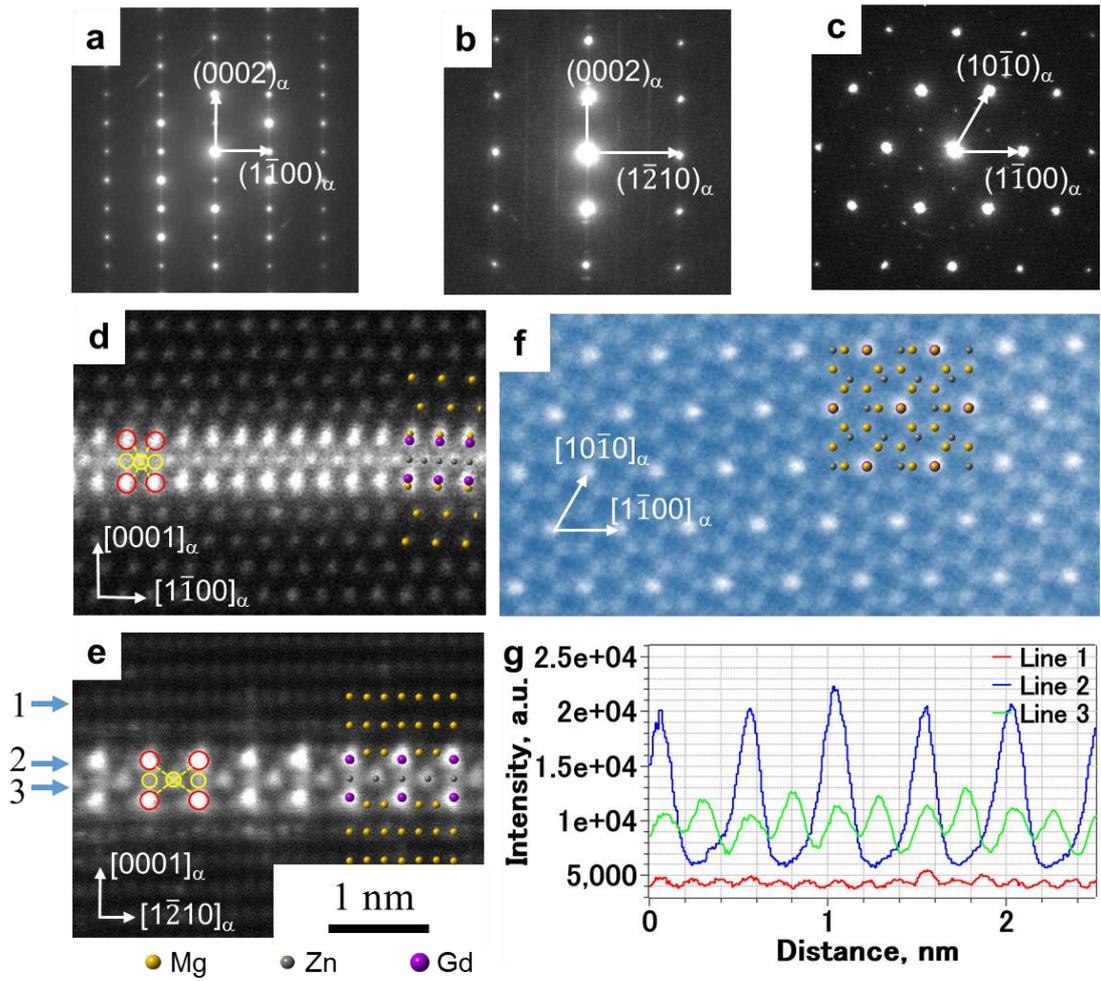

Figure 3. High-mag view of G.P. zone from different zone axes by HAADF-STEM. a-c) Diffraction patterns taken from [11-20]$_\alpha$, [10-10]$_\alpha$ and [0001]$_\alpha$, respectively. d) Atomic structure viewed from [11-20]$_\alpha$, e) Atomic structure viewed from [10-10]$_\alpha$, f) Atomic structure viewed from [0001]$_\alpha$. g) Intensity profile from row 1-3 indicated in e). In addition, the relaxed atomic structure is superimposed on d-f).

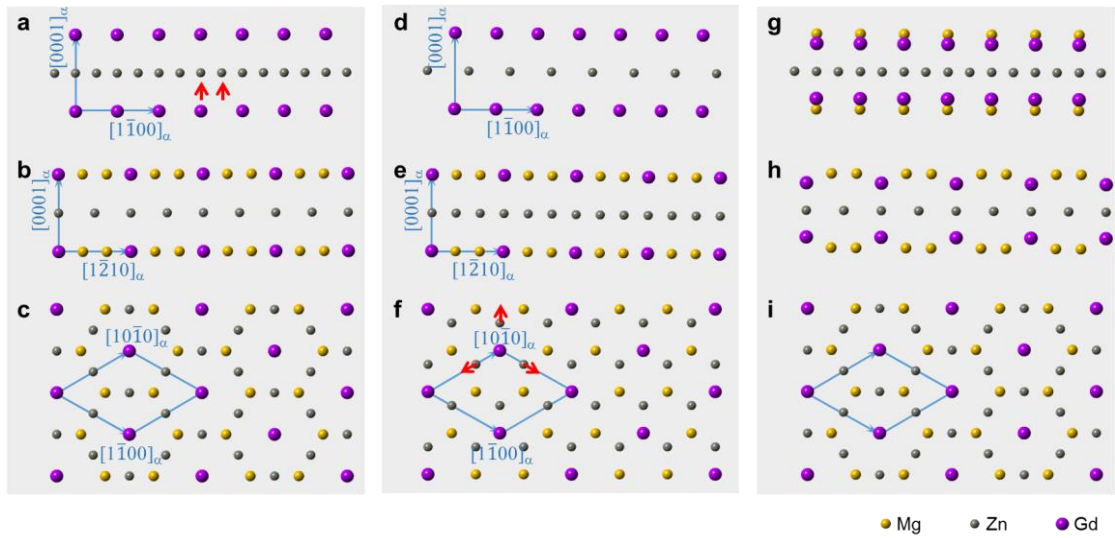

Figure 4. Atomic structures of G.P. zone in different models. a-c) Rigid atomic structure proposed in this study, d-f) Atomic structure model by Nie *et al.* [9], g-h) Relaxed atomic structure for a-c), respectively, by first-principles calculation. Figure a), d) and g) are viewed from [11-20]$_\alpha$, Figure b), e) and h) viewed from [10-10]$_\alpha$, Figure c), f), and i) are viewed from [0001]$_\alpha$.

**Supplementary materials**

Outline:

1. HAADF-STEM view of G.P. zone
2. Atomic positions in supercell before/after atomic relaxation

   2.1 Atomic positions for different rigid models proposed for G.P. zone

   2.2 Relaxed atomic structure

1. HAADF-STEM view of G.P. zone

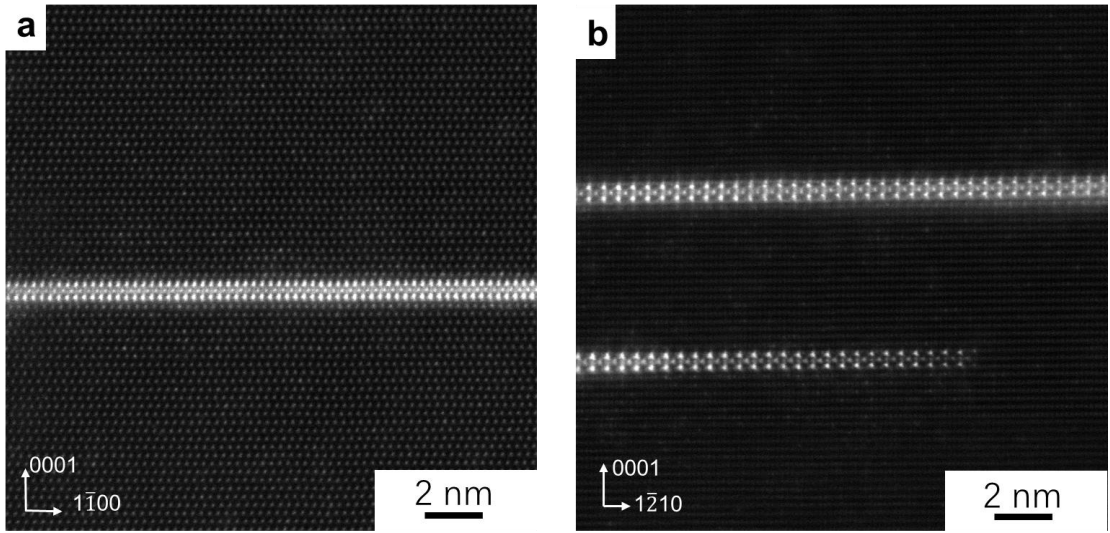

Figure S1. View of G.P. zone from different zone axes, a) [11-20]α, b)[10-10]α.

2. Atomic positions in supercell before/after relaxation

2.1 Atomic positions for different rigid models proposed for G.P. zone

1) Model 1: This study

| | | | |
|---|---|---|---|
| Mg | 0.00000 | 0.00000 | 0.00000 |
| Mg | 0.33333 | 0.33333 | 0.00000 |
| Mg | 0.66667 | 0.66667 | 0.00000 |
| Mg | 0.00000 | 0.33333 | 0.10000 |
| Mg | 0.66667 | 0.00000 | 0.10000 |
| Mg | 0.33333 | 0.66667 | 0.10000 |
| Mg | 0.00000 | 0.00000 | 0.20000 |
| Mg | 0.33333 | 0.33333 | 0.20000 |
| Mg | 0.66667 | 0.66667 | 0.20000 |
| Mg | 0.00000 | 0.33333 | 0.30000 |
| Mg | 0.66667 | 0.00000 | 0.30000 |
| Mg | 0.33333 | 0.66667 | 0.30000 |
| Gd | 0.00000 | 0.00000 | 0.40000 |
| Mg | 0.33333 | 0.33333 | 0.40000 |
| Mg | 0.66667 | 0.66667 | 0.40000 |
| Zn | 0.00000 | 0.50000 | 0.50000 |
| Zn | 0.50000 | 0.00000 | 0.50000 |
| Zn | 0.50000 | 0.50000 | 0.50000 |
| Gd | 0.00000 | 0.00000 | 0.60000 |
| Mg | 0.33333 | 0.33333 | 0.60000 |
| Mg | 0.66667 | 0.66667 | 0.60000 |
| Mg | 0.00000 | 0.33333 | 0.70000 |
| Mg | 0.66667 | 0.00000 | 0.70000 |
| Mg | 0.33333 | 0.66667 | 0.70000 |
| Mg | 0.00000 | 0.00000 | 0.80000 |
| Mg | 0.33333 | 0.33333 | 0.80000 |
| Mg | 0.66667 | 0.66667 | 0.80000 |
| Mg | 0.00000 | 0.33333 | 0.90000 |
| Mg | 0.66667 | 0.00000 | 0.90000 |
| Mg | 0.33333 | 0.66667 | 0.90000 |

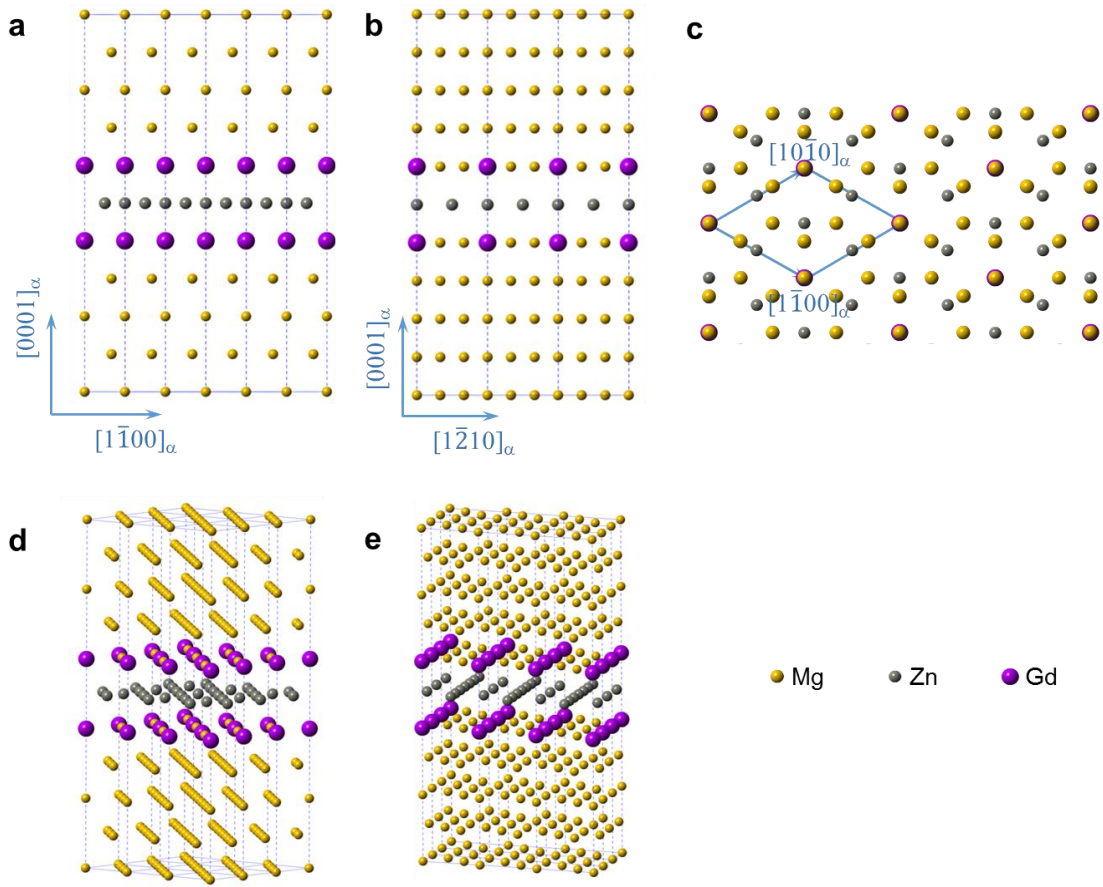

Figure S2. Atomic structure of Model 1 for G. P. Zone in different view. a) [11-20]$_\alpha$, b) [10-10]$_\alpha$, c) [0001]$_\alpha$, d) a little tilt from [11-20]$_\alpha$, e) a little tilt from [10-10]$_\alpha$

2) Model 2: Modified model by Nie *et al.* [9]

| | | | |
|---|---|---|---|
| Mg | 0.00000 | 0.00000 | 0.00000 |
| Mg | 0.33333 | 0.33333 | 0.00000 |
| Mg | 0.66667 | 0.66667 | 0.00000 |
| Mg | 0.00000 | 0.33333 | 0.10000 |
| Mg | 0.66667 | 0.00000 | 0.10000 |
| Mg | 0.33333 | 0.66667 | 0.10000 |
| Mg | 0.00000 | 0.00000 | 0.20000 |
| Mg | 0.33333 | 0.33333 | 0.20000 |
| Mg | 0.66667 | 0.66667 | 0.20000 |
| Mg | 0.00000 | 0.33333 | 0.30000 |
| Mg | 0.66667 | 0.00000 | 0.30000 |
| Mg | 0.33333 | 0.66667 | 0.30000 |
| Gd | 0.00000 | 0.00000 | 0.40000 |
| Mg | 0.33333 | 0.33333 | 0.40000 |
| Mg | 0.66667 | 0.66667 | 0.40000 |
| Zn | 0.00000 | 0.33333 | 0.50000 |
| Zn | 0.66667 | 0.00000 | 0.50000 |
| Zn | 0.33333 | 0.66667 | 0.50000 |
| Gd | 0.00000 | 0.00000 | 0.60000 |
| Mg | 0.33333 | 0.33333 | 0.60000 |
| Mg | 0.66667 | 0.66667 | 0.60000 |
| Mg | 0.00000 | 0.33333 | 0.70000 |
| Mg | 0.66667 | 0.00000 | 0.70000 |
| Mg | 0.33333 | 0.66667 | 0.70000 |
| Mg | 0.00000 | 0.00000 | 0.80000 |
| Mg | 0.33333 | 0.33333 | 0.80000 |
| Mg | 0.66667 | 0.66667 | 0.80000 |
| Mg | 0.00000 | 0.33333 | 0.90000 |
| Mg | 0.66667 | 0.00000 | 0.90000 |
| Mg | 0.33333 | 0.66667 | 0.90000 |

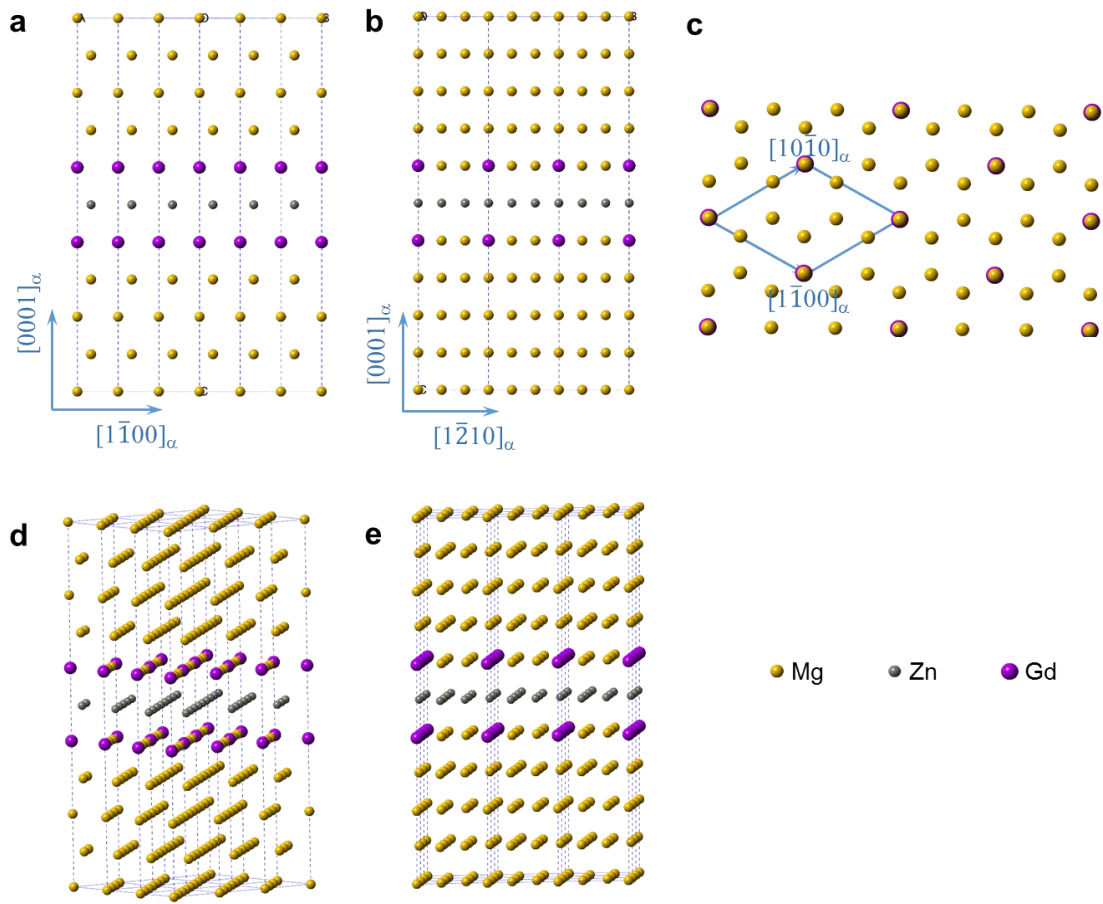

Figure S3. Atomic structure of Model 2 for G. P. Zone in different view. a) [11-20]$_\alpha$, b) [10-10]$_\alpha$, c) [0001]$_\alpha$, d) a little tilt from [11-20]$_\alpha$, e) a little tilt from [10-10]$_\alpha$

3) Model 3: Model proposed by *Li et al.* [16]

| | | | |
|---|---|---|---|
| Mg | 0.00000 | 0.00000 | 0.00000 |
| Mg | 0.33333 | 0.33333 | 0.00000 |
| Mg | 0.66667 | 0.66667 | 0.00000 |
| Mg | 0.00000 | 0.33333 | 0.10000 |
| Mg | 0.66667 | 0.00000 | 0.10000 |
| Mg | 0.33333 | 0.66667 | 0.10000 |
| Mg | 0.00000 | 0.00000 | 0.20000 |
| Mg | 0.33333 | 0.33333 | 0.20000 |
| Mg | 0.66667 | 0.66667 | 0.20000 |
| Mg | 0.00000 | 0.33333 | 0.30000 |
| Mg | 0.66667 | 0.00000 | 0.30000 |
| Mg | 0.33333 | 0.66667 | 0.30000 |
| Gd | 0.00000 | 0.00000 | 0.40000 |
| Mg | 0.33333 | 0.33333 | 0.40000 |
| Mg | 0.66667 | 0.66667 | 0.40000 |
| Zn | 0.58333 | 0.50000 | 0.50000 |
| Mg | 0.25000 | 0.16667 | 0.50000 |
| Mg | 0.91667 | 0.83333 | 0.50000 |
| Gd | 0.00000 | 0.00000 | 0.60000 |
| Mg | 0.33333 | 0.33333 | 0.60000 |
| Mg | 0.66667 | 0.66667 | 0.60000 |
| Mg | 0.00000 | 0.33333 | 0.70000 |
| Mg | 0.66667 | 0.00000 | 0.70000 |
| Mg | 0.33333 | 0.66667 | 0.70000 |
| Mg | 0.00000 | 0.00000 | 0.80000 |
| Mg | 0.33333 | 0.33333 | 0.80000 |
| Mg | 0.66667 | 0.66667 | 0.80000 |
| Mg | 0.00000 | 0.33333 | 0.90000 |
| Mg | 0.66667 | 0.00000 | 0.90000 |
| Mg | 0.33333 | 0.66667 | 0.90000 |

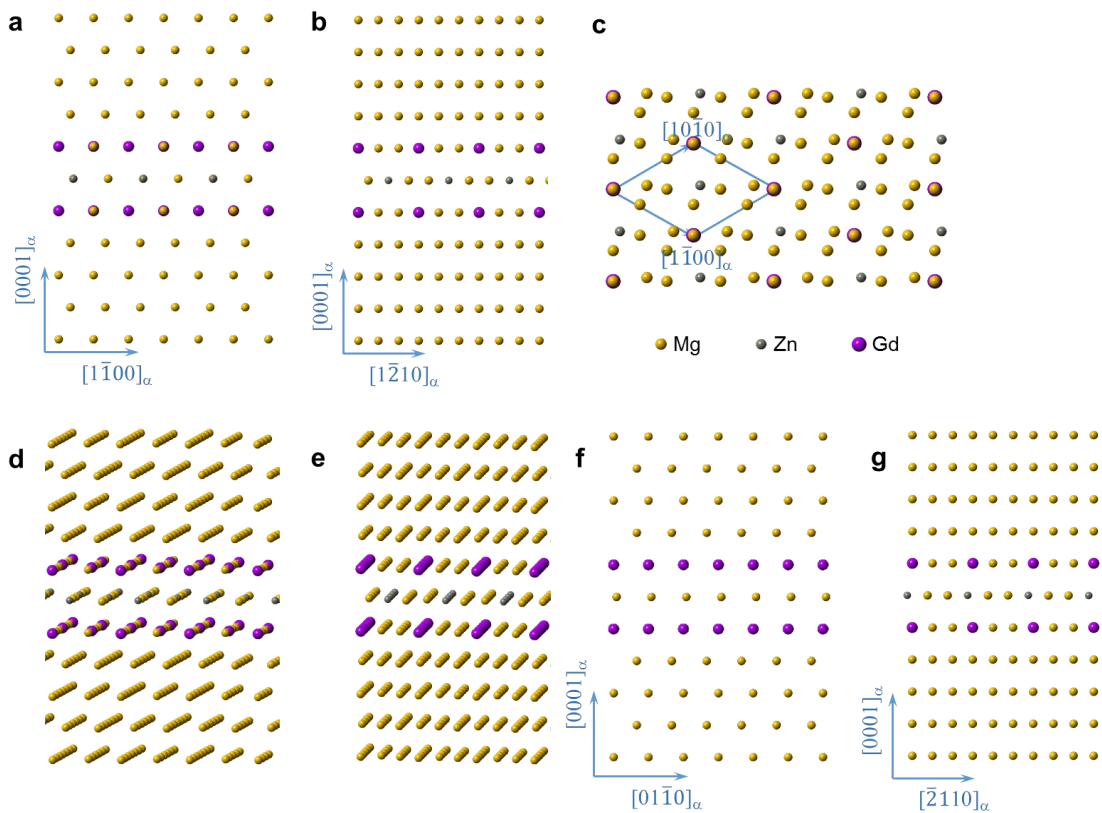

Figure S4. Atomic structure of Model 3 for G. P. Zone in different view. a) [11-20]$_\alpha$, b) [10-10]$_\alpha$, c) [0001]$_\alpha$, d) a little tilt from [11-20]$_\alpha$, e) a little tilt from [10-10]$_\alpha$, f) [-2110]$_\alpha$, g) [01-10]$_\alpha$

2.2 Relaxed atomic structure

1) Relaxed structure of Model 1 or 2

| | | | |
|---|---|---|---|
| Mg | -0.00006 | 0.00006 | 0.000362 |
| Mg | 0.33327 | 0.33338 | 0.0004 |
| Mg | 0.66662 | 0.66673 | 0.0004 |
| Mg | -0.00011 | 0.33412 | 0.103166 |
| Mg | 0.66588 | 0.00011 | 0.103166 |
| Mg | 0.33392 | 0.66608 | 0.103164 |
| Mg | 0.00001 | -0.00001 | 0.204466 |
| Mg | 0.33334 | 0.33333 | 0.207212 |
| Mg | 0.66667 | 0.66666 | 0.207212 |
| Mg | 0.00004 | 0.32824 | 0.308455 |
| Mg | 0.67176 | -0.00004 | 0.308455 |
| Mg | 0.32835 | 0.67165 | 0.308454 |
| Mg | 0.33339 | 0.33328 | 0.406308 |
| Mg | 0.66672 | 0.66661 | 0.406308 |
| Mg | 0.33337 | 0.33329 | 0.593515 |
| Mg | 0.66671 | 0.66663 | 0.593515 |
| Mg | 0.00003 | 0.32790 | 0.691603 |
| Mg | 0.67210 | -0.00003 | 0.691603 |
| Mg | 0.32801 | 0.67199 | 0.691606 |
| Mg | 0.00001 | -0.00001 | 0.795791 |
| Mg | 0.33334 | 0.33333 | 0.793148 |
| Mg | 0.66667 | 0.66666 | 0.793148 |
| Mg | -0.00011 | 0.33400 | 0.897327 |
| Mg | 0.66600 | 0.00011 | 0.897327 |
| Mg | 0.33380 | 0.66620 | 0.897326 |
| Zn | 0.00007 | 0.49865 | 0.499749 |
| Zn | 0.50135 | -0.00007 | 0.499749 |
| Zn | 0.49883 | 0.50117 | 0.499749 |
| Gd | 0.00001 | -0.00001 | 0.430585 |
| Gd | 0.00001 | -0.00001 | 0.566732 |

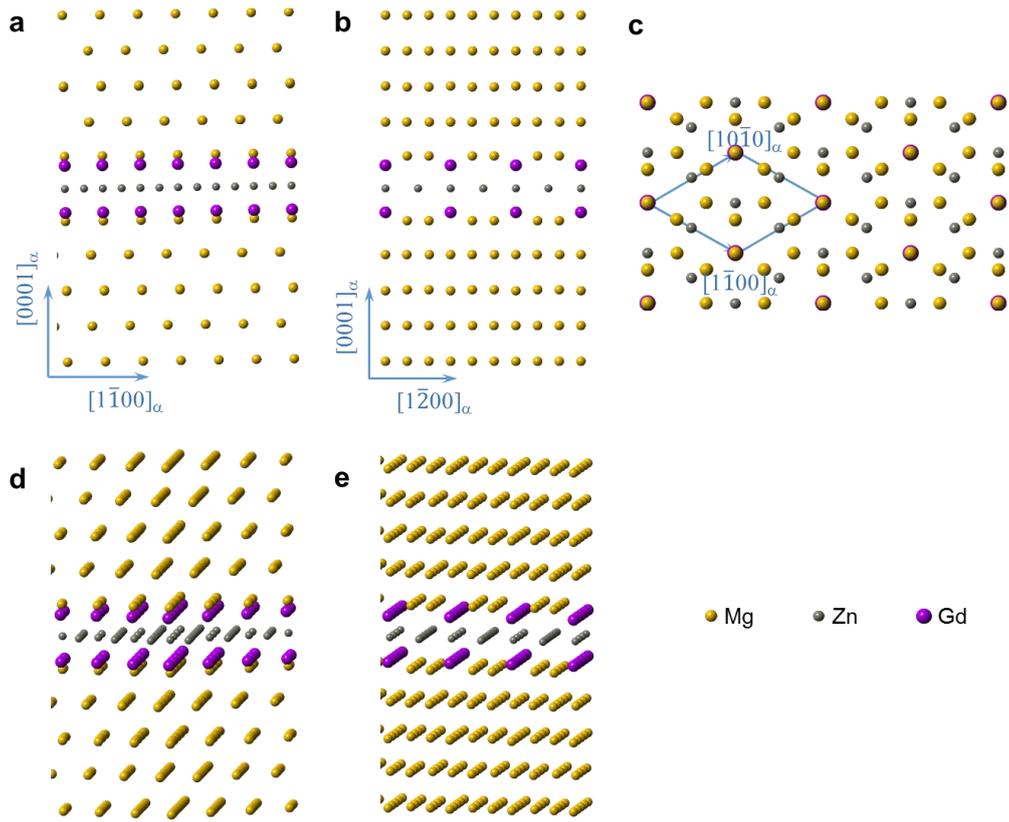

Figure S5. Relaxed atomic structure of Model 1 or 2 for G. P. Zone in different view. a) [11-20]$_\alpha$, b) [10-10]$_\alpha$, c) [0001]$_\alpha$, d) a little tilt from [11-20]$_\alpha$, e) a little tilt from [10-10]$_\alpha$

2) Relaxed structure of Model 3

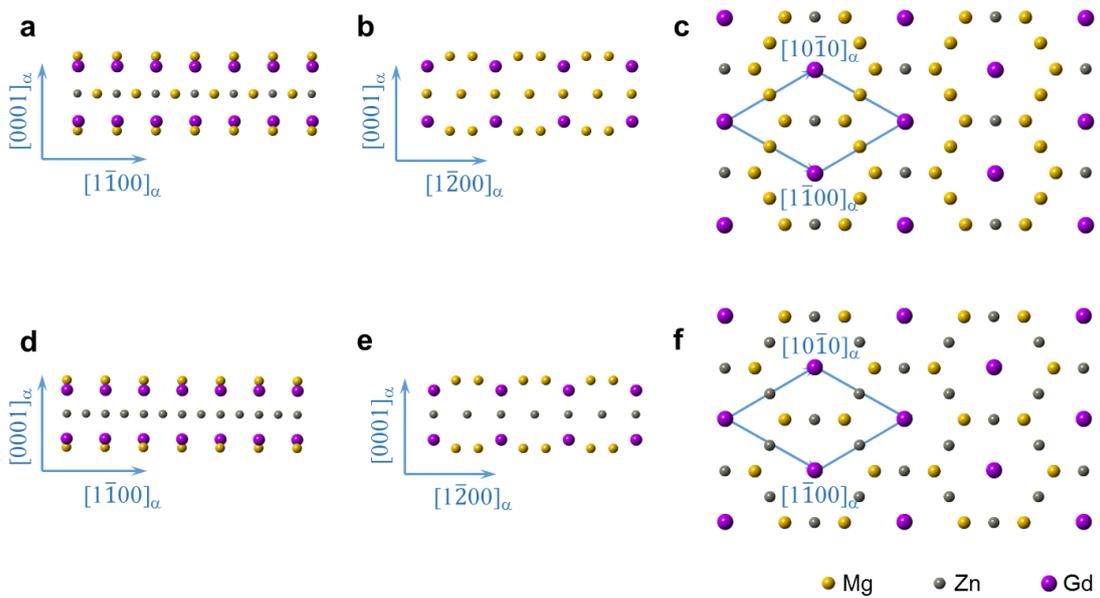

Figure S6. Relaxed atomic structure of Model 3 for G. P. Zone in different view. a-c)

original model, d-f) modified model with only Zn atoms in middle layer. a,d) [11-20]$_\alpha$, b,e) [10-10]$_\alpha$, c,f) [0001]$_\alpha$,